\documentclass[review]{elsarticle}
\usepackage{graphicx}
\usepackage{amsmath}   
\usepackage{newtxtext,newtxmath}

\begin{document}

\begin{frontmatter}

\title{A simple model for multiple coincidence SiPM dark noise }

\author[infnroma1]{Valerio ~Bocci\corref{cor}}
\ead{valerio.bocci@roma1.infn.it}

\cortext[cor]{Corresponding author}
\address[infnroma1]{INFN Sezione di Roma, Rome, Italy}
\begin{abstract}
Silicon photomultipliers are photon sensors based on an array of pixels, each consisting of a single photon avalanche photodiode (SPAD). 
Reading of the array is not done on a single-pixel basis, but all pixels are read simultaneously.
Because of thermal agitation and other phenomena, each pixel can fire even in the dark with a small but not zero probability of adding to the output signal a fixed current.
Each pixel has an average random event rate, and the dark noise rate is due to the sum of any pixel contribution. 
Statistically, two or more pixels may fire simultaneously, giving a current signal equal to the sum of the pixels involved.
This eventuality has a lower probability of a more significant number of fired pixels.
This paper aims to find simple formulas that relate dark noise and the rate of having multiple multiplicity events.

\end{abstract}

\begin{keyword}
rate formula  \sep SiPM \sep all-in-one detector

\end{keyword}

\end{frontmatter}

\section{Introduction}

\begin{figure}
\centering
\includegraphics[width=0.99\linewidth]{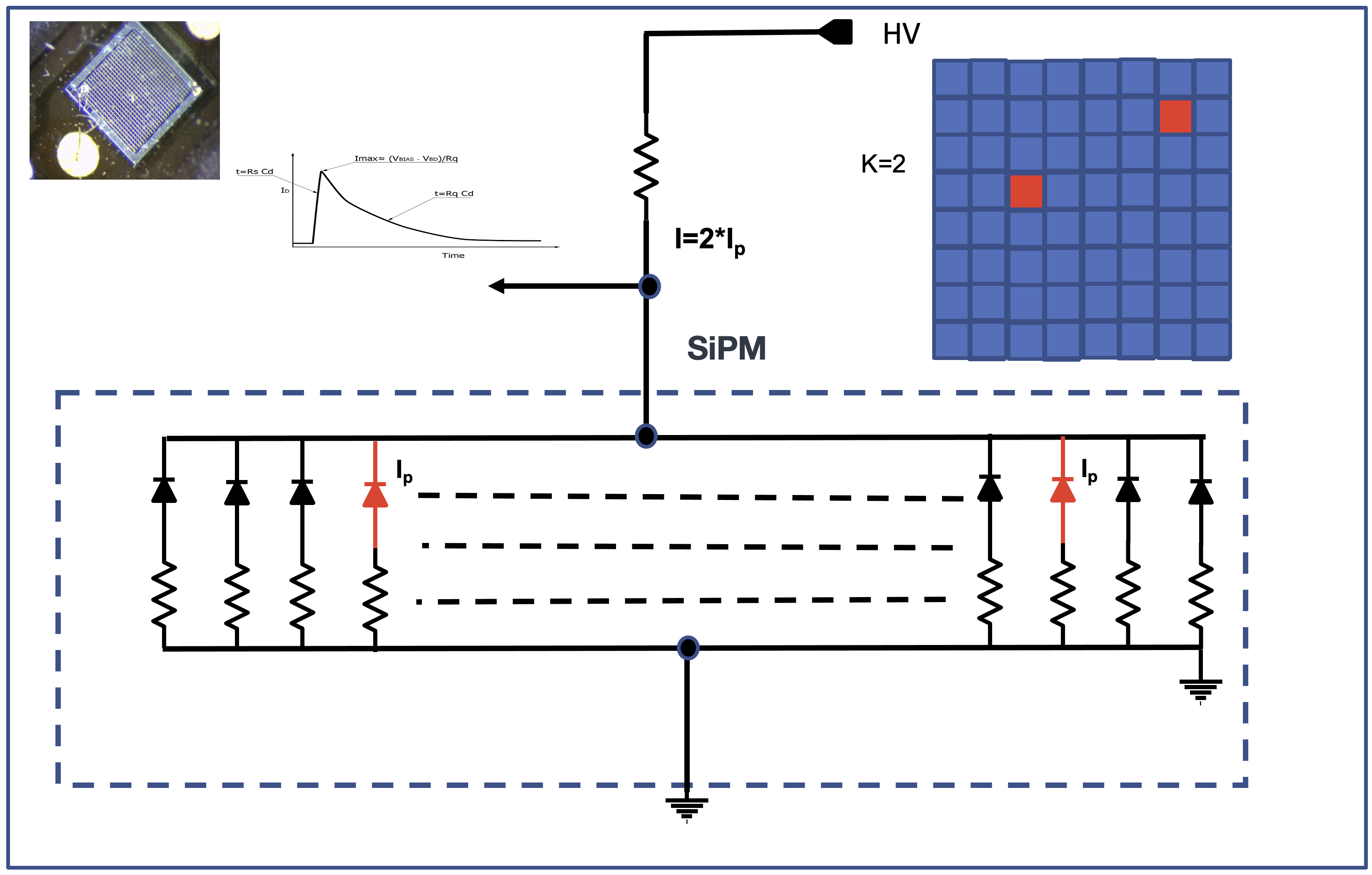}
\caption{The SiPM generic architecture}
\label{fig:SiPM}
\end{figure}

Silicon photomultipliers are photon sensors developed in the late 1980s\cite {Gasanov:1988} and commercially distributed in the early 2000s, and they have reached technological maturity in the past decade. 
Their use ranges from high-energy physics used as light detectors in scintillation and Cherenkov particle detectors as sensors for bioluminescence to the automotive market as Lidar. Although their use has not replaced photomultipliers, their nature as solid-state sensors makes them attractive in multiple areas.
As shown schematically in Figure \ref{fig:SiPM}
SIPM is based on an array of pixels, each consisting of a single photon avalanche photodiode (SPAD). 
Each SPAD works in Geiger mode so that when a photon hits a SPAD, it produces a discharge that is stopped by a quenching resistor.
One head of each SiPMs quenching resistor is electrically connected to the same potential, and then the output is the sum of all the pixels.
Because of thermal agitation and other phenomena, each pixel can fire even in the dark with a small but not zero probability of adding to the output signal a fixed current.
Statistically, two or more pixels may fire simultaneously, giving a current signal equal to the sum of the pixels involved.
This eventuality has a lower probability of a more significant number of fired pixels.
This paper aims to find simple formulas that relate dark noise and the rate of having multiple multiplicity events. 

\section{The accidental coincidence formula}

The advent of Muller Geiger counters in the early 1930s posed the problem of reducing these devices' accidental coincidences (noise). The noise rate of a single Geiger Muller counter could be reduced by using two or more counters in coincidence. This approach does not eliminate the random effects but dramatically reduces the rate. We find a first formulation of the rate calculation of two-fold accidental coincidences in a 1938 paper by Carl Eckart \cite{AC_Eckart:1938}, and a correction for n-fold generalization comes out from Janossy in \cite {AC_Janossy1:1944} and if we don't consider as noted from \cite {AC_Janossy2:1944} at the Schroedinger  article about this item  \cite {AC_SCHROEDINGER:1944} of $t^n$ contribution the general formula for the accidental coincidence can be written as

$$
f_{AC}(n)=\prod_{i=1}^n \boldsymbol{f}_i \tau_i \sum_{i=1}^n \frac{1}{\tau_i}
$$

If all the pulse have equal widths $\tau _i =\tau $ and equal rates 
$f_i=f$ the equation can  reduced to:

\begin{equation}
f_{k}=k \tau^{k-1} f_{detector}^{k}
\label{eq:generic_coinc_eq}
\end{equation}
 where we require k events of k detector in coincidence.
\section{The accidental coincidence formula for an array N pixel }
In the case of SiPM we have an array of N pixel that can fire with a given mean noise rate.  If we want to use the  formula \eqref {eq:generic_coinc_eq} for k pixel fired in coincidence in an array of N pixel the formula above should be multiplied by the number of combinations of k pixel in the N pixels of the array:

\begin{equation} f_{k}=k \tau^{k-1} f^{k} N _{Comb} \end{equation}

The number of combination of k objects in N objects is given by the binomial

\begin{equation}N_{Comb}=\binom{N}{k} =\frac{N !}{k !(N-k) !} \end{equation}

The number of pixel in a SiPM array is normally of the order of thousand instead the coincidence pixel that we want to evaluate is some units than we have $  k \ll N $ and we can  use the Stirling approximation for the binomial:

\begin{equation}N_{Comb}=\frac{N !}{k !(N-k) !} \sim \frac{N^{k}}{k !}\end{equation}

Therefore, the frequency of k pixels of N pixels each with rate f in coincidence comes:

\begin{equation}
f_{coinc}=k \tau^{k-1} f_{pixel}
^k \frac {N^k}{k!}
\label{eq:coinc_formula}
\end{equation}
This formula presupposes knowledge of the dark frequency of each individual pixel $f_{pixel}$, a frequency not directly measurable. Therefore, we seek a method to estimate the frequency of a single pixel.
\section{Relative Scale of Coincidence Rates to Total Dark Noise Estimation }
An important parameter that characterizes the SiPM is the dark count rate, which is the noise rate with a threshold of about 0.5 photoelectrons. This corresponds to the sum of all the frequencies of coincidences
\begin{equation}
f_{dark}=\sum_{k=1}^{N} f_k
\end{equation}

We can use Equation (\ref{eq:coinc_formula})  to understand the contribution of each term to this summation by calculating the ratio between the coincidence rates $k$ and $k-1$.

\begin{equation}
\frac {f_{k}}{f_{k-1}}=
\frac
{k \tau^{k-1} f_{pixel}^k \frac {N^k}{k!}}
{(k-1) \tau^{k-2} f_{pixel}^{(k-1)} \frac {N^{(k-1)}}{(k-1)!}}=
\frac
{k \tau f_{pixel} {N}}
{(k-1) {k}}=
\frac
{\tau f_{pixel} {N}}
{(k-1) }=
\frac
{\tau }
{ \tau_{pixel}}
\frac
{{N}}
{(k-1)}
\end{equation}

we obtain the formula :
\begin{equation}
\frac {f_{k}}{f_{k-1}}=
\frac
{\tau }
{ \tau_{pixel}}
\frac
{{N}}
{(k-1)}
\label{eq:coinc_ratio_formula}
\end{equation}
As an example, if If we consider commonly found values for some SiPMs as $\tau=10ns$, $f_{pixel}=100 Hz $,  $\tau_{pixel}$ with a sensor of  $N=1000$ pixels

We have  $\frac
{\tau }
{ \tau_{pixel}}=10^{-5}$

\begin{equation}
\frac {f_{k}}{f_{k-1}}=
\frac
{10^{-2}}
{(k-1)}
\end{equation}
To obtain the order of magnitude of the ratio, simply take the $\log_{10}$

\begin{equation}
\log_{10}{\frac {f_{k}}{f_{k-1}}}=
\log_{10} {\frac
{10^{-2}}
{(k-1)}}
\end{equation}
the ratio between $f_2$ and $f_1$ is the order of two order of magnitude and the ratio from $f_3$ and $f_2$ is other 2.3 order of magnitude.
The practical impact of these calculations is that the dark noise is almost entirely composed of single coincidences, so it is reasonable to approximate it to $f_1$

\begin{equation}
f_{Dark} \approx f_{1}= \tau^{0} f_{pixel}
^1 \frac {N^1}{1!}=f_{pixel}{N}
\end{equation}

from which we have:

\begin{equation}
\boxed{f_{pixel}=\frac {f_{Dark}}{N}}
\end{equation}
This approximation for \( f_{\text{pixel}} \) is of practical significance, as it only requires the measurement of \( f_{\text{dark}} \). This can be achieved by placing the SiPM in darkness and using a signal discriminator with a threshold around 0.5 photons, permitting the measurement of pulse frequency using an hardware like \cite {ArduSiPM1:2014}.
By substituting the expression into formula \eqref {eq:coinc_ratio_formula}, the formula for the rate of multiple coincidences for a generic SiPM is obtained, once its $f_{\text{dark}}$ and its $\tau$ are known.

\begin{equation}
\boxed{
f_{\text{coinc}_k}=\tau^{k-1} \frac{f_{\text{dark}}^k}{(k-1)!} = 
f_{\text{dark}} \frac{(\tau f_{\text{dark}})^{k-1}}{(k-1)!}
}
\footnote{Formula for multiple coincidences in a SiPM proposed in this paper.}
\label{eq:VB_EQ}
\end{equation}

\section{Conclusions}
The equation referenced as \eqref{eq:VB_EQ} provides a reliable approximation for the rate of k-fold coincidences in pixels activated within an N-pixel matrix of a SiPM detector. This approach is also applicable in similar contexts. 
In this formula, it is assumed that all pixels in the SiPM have identical characteristics.


\begin{thebibliography}{10}
\expandafter\ifx\csname url\endcsname\relax
  \def\url#1{\texttt{#1}}\fi
\expandafter\ifx\csname urlprefix\endcsname\relax\def\urlprefix{URL }\fi


\bibitem{AC_Eckart:1938}
Accidental Coincidences in Counter Circuits.
Carl Eckart and Francis R. Shonka.
Phys. Rev. 53, 752 – Published 1 May 1938
doi: 10.1103/PhysRev.53.752


\bibitem{AC_Janossy1:1944}  Rate of n-fold Accidental Coincidences. Janossy, L. Nature 153, 165 (1944). doi: 10.1038/153165a0

\bibitem{AC_SCHROEDINGER:1944} Rate of n-fold Accidental Coincidences.
Erwin Schroedinger
Nature volume 153, pages592–593 (1944)
doi: 10.1038/153592b0 

\bibitem {AC_Janossy2:1944} JANOSSY, L. Rate of n-fold Accidental Coincidences. Nature 153, 593 (1944). https://doi.org/10.1038/153593a0

\bibitem {Gasanov:1988} Gasanov A. G. et al. 1988 Soviet Technical. Physics Letters Sov. 14, 313 (1988)

\bibitem{ArduSiPM1:2014}
V. Bocci, G. Chiodi, F. Iacoangeli, M. Nuccetelli and L. Recchia, "The ArduSiPM a compact trasportable Software/Hardware Data Acquisition system for SiPM detector," 2014 IEEE Nuclear Science Symposium and Medical Imaging Conference (NSS/MIC), 2014, pp. 1-5, doi: 10.1109/NSSMIC.2014.7431252.




\end{thebibliography}
\end{document}